\documentclass[fleqn,usenatbib]{aa}  

\usepackage{graphicx}
\usepackage{txfonts}

\usepackage{threeparttable}
\begin{document} 

   \title{Gravitational collapse and accretion flows in the hub filament system G323.46-0.08}


   \author{Yingxiu Ma
          \inst{1,2}
          \and
          Jianjun Zhou\inst{1,3,4}
          \and
          Jarken Esimbek\inst{1,3,4}
          \and
          Willem Baan\inst{1,5}
          \and
          Dalei Li\inst{1,2,3,4}
          \and
          Xindi Tang\inst{1,2,3,4} 
          \and
          Yuxin He\inst{1,2,3,4}
          \and
          Weiguang Ji\inst{1}          
          \and
          Dongdong Zhou\inst{1}
          \and
          Gang Wu\inst{6} 
          \and
          Kadirya Tursun\inst{1} 
          \and
          Toktarkhan Komesh\inst{7,8} 
          }
             
   \institute{XingJiang Astronomical Observatory, Chinese Academy of Sciences,
             Urumqi 830011, PR China\\
              \email{zhoujj@xao.ac.cn; jarken@xao.ac.cn; mayingxiu@xao.ac.cn}
         \and
             University of Chinese Academy of Sciences, Beijing 100049, PR China
         \and
             Key Laboratory of Radio Astronomy, Chinese Academy of Sciences, Urumqi 830011, PR China
         \and
             Xinjiang Key Laboratory of Radio Astrophysics, Urumqi 830011, PR China
        \and
             Netherlands Institute for Radio Astronomy, ASTRON, 7991 PD Dwingeloo, The Netherlands
         \and
            Max-Planck-Institut f\"{u}r Radioastronomie, Auf dem H\"{u}gel 69, D-53121 Bonn, Germany
         \and
           Energetic Cosmos Laboratory, Nazarbayev University, Astana 010000, Kazakhstan
         \and
           Faculty of Physics and Technology, Al-Farabi Kazakh National University, Almaty, 050040, Kazakhstan
             }    


 
 \abstract
{We studied the hub-filament system G323.46-0.08 based on archival molecular line data from the SEDIGISM $^{13}$CO survey and infrared data from the GLIMPSE, MIPS, and Hi-GAL surveys. G323.46-0.08 consists of three filaments, F-north, F-west, and F-south, that converge toward the central high-mass clump AGAL\,323.459-0.079. F-west and Part\,1 of the F-south show clear large-scale velocity gradients 0.28 and 0.44\,km s$^{-1}$\,pc$^{-1}$, respectively. They seem to be channeling materials into AGAL\,323.459-0.079. The minimum accretion rate was estimated to be 1216\,M$_\odot$\,Myr$^{-1}$. A characteristic V-shape appears around AGAL\,323.459-0.079 in the PV diagram, which traces the accelerated gas motions under gravitational collapse. This has also been supported by model fitting results.  All three filaments are supercritical and they have fragmented into many dense clumps. The seesaw patterns near most dense clumps in the PV diagram suggests that mass accretion also occurs along the filament toward the clumps.
 Our results show that filamentary accretion flows appear to be an important mechanism for supplying the materials necessary to form the central high-mass clump AGAL\,323.459-0.079 and to propel the star forming activity taking place therein.
}
   
   \maketitle
%

\section{Introduction}

Recent observations have shown that filaments are ubiquitous in molecular clouds \citep{Molinari2010a,Li2016, Mattern2018, Schisano2020}. 
Most dense clumps and cores are formed in filaments and they may play a key role in the star formation process \citep{Andre2013}. 
The filaments in molecular clouds may overlap to form a web, creating junctions of filaments and the hub-filament systems \citep{Myers2009}. 
In a scenario of global hierarchical collapse of molecular clouds dominated by gravity, these filaments constitute channels for gas funneled from the extended cloud to the dense clumps \citep{gomez2014,vazquez2019}. 
The clumps at the junction will accrete more materials from the surroundings and they become more massive and more likely to form high-mass star clusters. 
Compared to isotropic accretion, the accretion flow along the filament is a more efficient mechanism because it takes a longer time to gather mass for the high-mass star forming cores before channels are destroyed \citep{Myers2009}. 
Several hub-filament systems show evidences of channeling gas feeding to the junctions with ongoing high-mass star formation \citep{Kirk2013, Peretto2013, Peretto2014, Yuan2018}

For clouds in gravitational collapse, the gas flows will be accelerated towards the center of the potential well because of gravitational attraction \citep{gomez2014}. 
It is expected to produce a characteristic V-shape on the gas velocity structure, which traces the accelerated gas motion  around the central high mass clumps or cores. 
Such a feature has been detected within OMC-1 cloud \citep{hacar2017}. 

The molecular clouds in the region ($323^{\circ} <l < 324^{\circ}$, $ \left| b \right| < 0.5^{\circ}$) have been studied by \citet{Burton2013} as a sample of the Mopra Southern Galactic Plane CO Survey. 
Several select  sub-regions have been identified and labeled A through F. Their physical parameters were derived but without studying the clouds' structure and kinematics. However, the high angular resolution $^{13}$CO\,($J$\,=\,2--1) data from the Structure, Excitation and Dynamics of the Inner Galactic Interstellar Medium (SEDIGISM, \citet{Schuller2021}) survey reveals that molecular clouds in the region have very interesting structures and kinematics. 
Using this new $^{13}$CO\,($J$\,=\,2--1) survey data, we found that molecular cloud complex in sub-region F is a good cloud-cloud collision candidate \citep{ma2022}. We also noted that molecular clouds in the sub-region A may make up a hub-filament system (G323.46-0.08), showing evidence of accretion flow in filaments and gravitationally collapse in the junction region. This provides a good target for studying of the role of filaments in the evolution of the molecular cloud and ongoing star formation. \citet{Burton2013}  estimated a kinematic distance of 4.8\,kpc, and a column density (N$_{\rm\,H_2}$) of 8\,$\times$\,10$^{21}$\,cm$^{-2}$ for sub-region A.

In this paper, we present a detailed study of the hub-filament system G323.46-0.08. 
We describe the data in Section\,2 and present the physical parameters and kinematics of the G323.46-0.08 in Section\,3. 
Section\,4 discusses the evidences of convergence, accelerations, and young stellar objects (YSOs). 
Finally, Section\,5 presents the conclusions of this work.  

\section{Archival data}

\subsection{The $^{13}$CO\,($J$\,=\,2--1) emission data}

The $^{13}$CO\,($J$\,=\,2--1) data from the SEDIGISM survey\footnote{The SEDIGISM survey data have been obtained with the APEX telescope under the programs F-9315(A) and 193.C-0584(A). APEX is a collaboration between the Max-Planck-Institut f\"ur Radioastronomie, the European Southern Observatory, and the Onsala Space Observatory. The processed data products are available from the SEDIGISM survey database located at https://sedigism.mpifr-bonn.mpg.de/index.html, which was constructed by James Urquhart and is hosted by the Max-Planck-Institut f\"ur Radioastronomy.} has been obtained the Atacama Pathfinder EXperiment (APEX) and has been used for kinematic analysis \citep{Schuller2021}. 
This survey mapped 84\,deg$^2$ of the Galactic plane ($-60^{\circ} <l < 31^{\circ}$, $ \left| b \right| < 0.5^{\circ}$) in several molecular transitions. 
The angular resolution of SEDIGISM is $\sim$\,$30''$, and the 1\,$\sigma$ sensitivity is about 0.8--1.0\,K for a 0.25\,km\,s$^{-1}$ channel width. 

\subsection{The infrared data}
Mid-infrared emission data at 3.6, 4.5, 5.8, and 8\,$\mu$m have been obtained from the Galactic Legacy Infrared Mid-Plane Survey Extraordinaire (GLIMPSE) \citep{Benjamin2003}. 
The 5\,$\sigma$ sensitivities at the four bands are 0.2, 0.2, 0.4, and 0.4\,mJy, respectively. The corresponding angular resolutions are between $1.5''$ and $1.9''$ \citep{Fazio2004}. 
We also used 24 and 70\,$\mu$m image data from the Multiband Infrared Photometer of the Spitzer MIPS Galactic Plane Survey (MIPSGAL) \citep{Carey2009}. 
The 5\,$\sigma$ sensitivity is 1.7\,mJy, and the corresponding resolutions are $6''$ and $18''$, respectively \citep{Rieke2004}.
 
The Herschel Infrared Galactic Plane Survey (Hi-GAL) data has been used to derive the dust temperature and column density distribution \citep{Molinari2010}). 
The angular resolutions of these Herschel maps are approximately $10.2''$, $13.5''$, $18.1''$, $25.0''$, and $36.4''$ at 70, 160, 250, 350, and 500\,$\mu$m, respectively. 

\subsection{Source catalogs}
The APEX Telescope Large Area Survey of the Galaxy (ATLASGAL) dense clump catalog \citep{Urquhart2018} and the Herschel Hi-GAL clump catalog \citep{Elia2017} were used to trace dense clumps and the embedded pre-stellar and proto-stellar cores. 
The GLIMPSE Point-Source catalog (GPSC), MIPSGAL \citep{Gutermuth2015} and 2MASS catalog \citep{Skrutskie2006} are used to trace the population of Class I and II YSOs .

\section{Results}

\subsection{Hub-filament system G323.46-0.08}

  \begin{figure*}
  \centering
  \includegraphics[trim={0.5cm 0cm 1.5cm 0.5cm},clip,width=18.5cm]{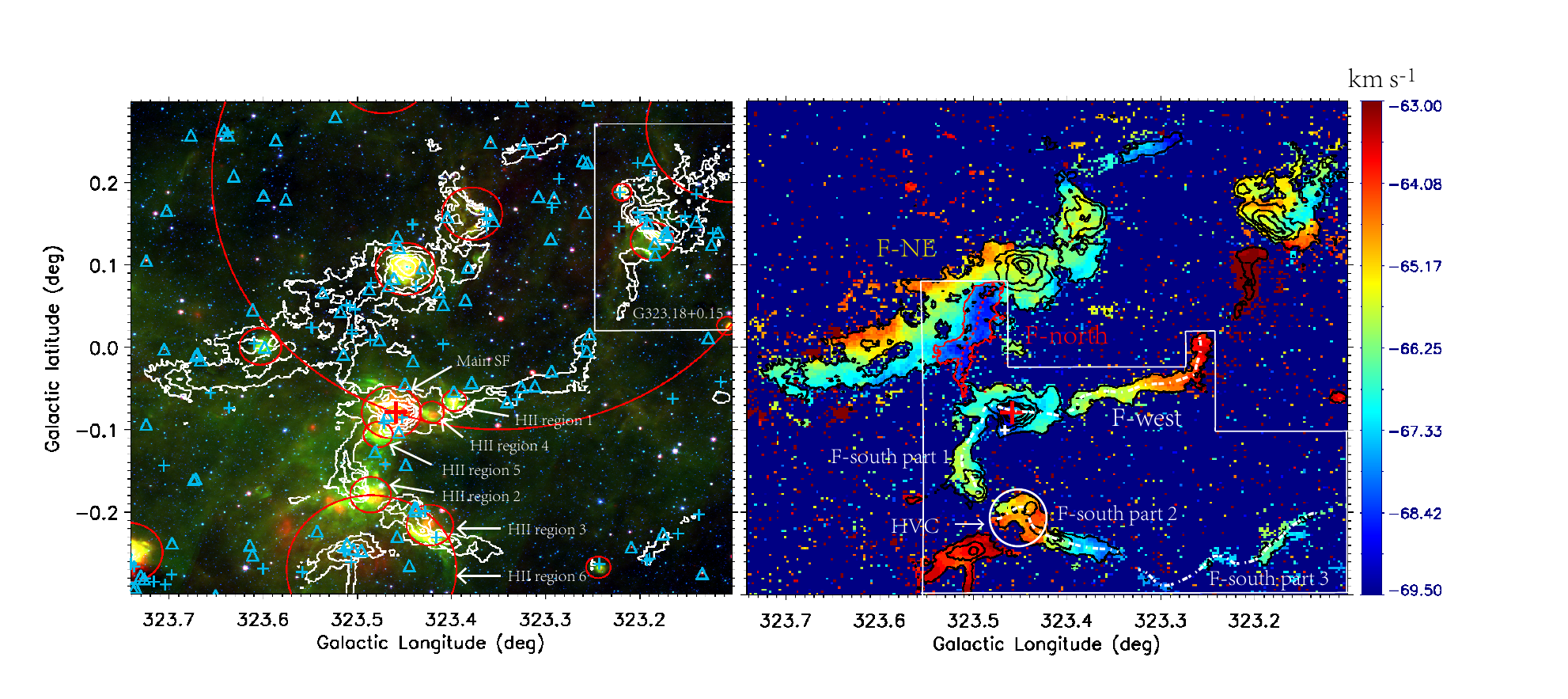}
 \caption{Infrared and molecular spectrum line emissions of G323.46-0.08. 
 Left: Three-color map of the G323.46-0.08 region. 
 Red, green, and blue backgrounds show the $24\,\mu$m, $8\,\mu$m, and $4.5\,\mu$m emission, respectively. 
The white contours denote the $^{13}$CO\,($J$\,=\,2--1) emission integrated velocity from -70 to -63\,km\,s$^{-1}$, which starts from 3\,K\,km\,s$^{-1}$ and with a step of 9\,K\,km\,s$^{-1}$. 
The cyan crosses and triangles are Class\,I and II YSOs, respectively. 
The red circles mark the H{\scriptsize II} regions identified by \citet{Anderson2014}. 
The red cross is the central high mass clump AGAL323.459-0.079 in hub-filament system G323.46-0.08. 
Right: Velocity map of G323.46-0.08. 
The color background is the velocity field of $^{13}$CO\,($J$\,=\,2--1). 
The black contours are the integrated intensity of the $^{13}$CO\,($J$\,=\,2--1) emission (same as in the left panel). The red contour marks the $^{13}$CO\,($J$\,=\,2--1) emission in the region of F-north integrated velocity from -70 to -68.25\,km\,s$^{-1}$.
The red dotted line and the white interrupted dashed line represent the ridgeline of F-north, F-west, and F-south. 
All ridge lines are formed by connecting the peak intensity points of $^{13}$CO\,($J$\,=\,2--1) in the filament). The white circle denotes the high velocity component appears in F-south part\,2.
              }
         \label{fig1}
  \end{figure*}
 
  \begin{figure*}
  \begin{centering}
  \includegraphics[trim={0.5cm 0cm 2cm 1cm},clip,width=19cm]{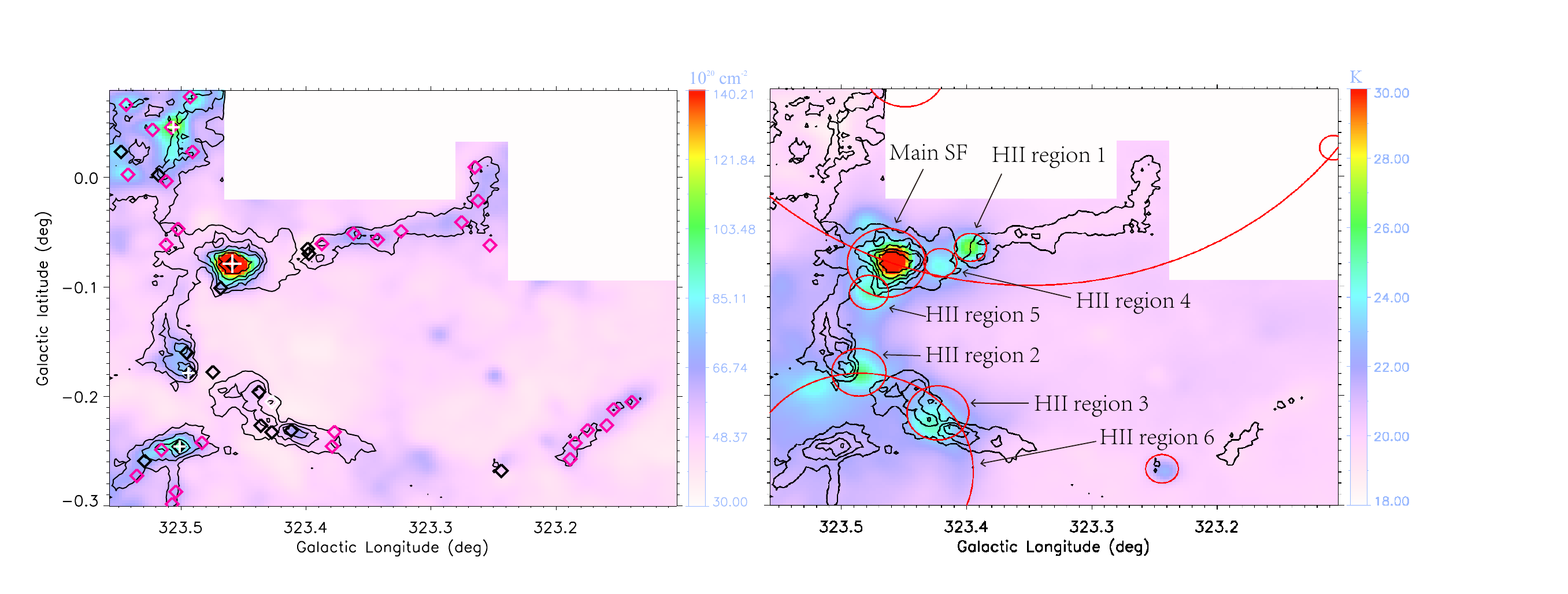}
\caption{H$_{2}$ column density (left) and dust temperature (right) distributions within the G323.46-0.08. 
The black contours indicate the $^{13}$CO\,($J$\,=\,2--1) integrated intensity in areas, where $T_{\rm mb}$ is higher than 3\,$\sigma$ (3\,K), 
The contour levels start from 3\,K\,km\,s$^{-1}$ and with a step of 9\,K\,km\,s$^{-1}$. 
The velocity interval of the integration is from -70 to -63\,km\,s$^{-1}$. 
The white, rose, and black diamonds are Herschel clumps, and represent different evolution stages of starless cores, prestellar, and protostellar, respectively \citep{Elia2017}. 
The white crosses are ATLASGAL 870 $\mu$m clumps \citep{Urquhart2018}. 
The red circles show the H{\scriptsize II} regions identified by \citet{Anderson2014}.}
         \label{fig3_appendix}
    \end{centering}     
  \end{figure*}

  \begin{figure}
  \centering
  \includegraphics[trim={1cm 0cm 0cm 1.5cm},clip,width=10cm]{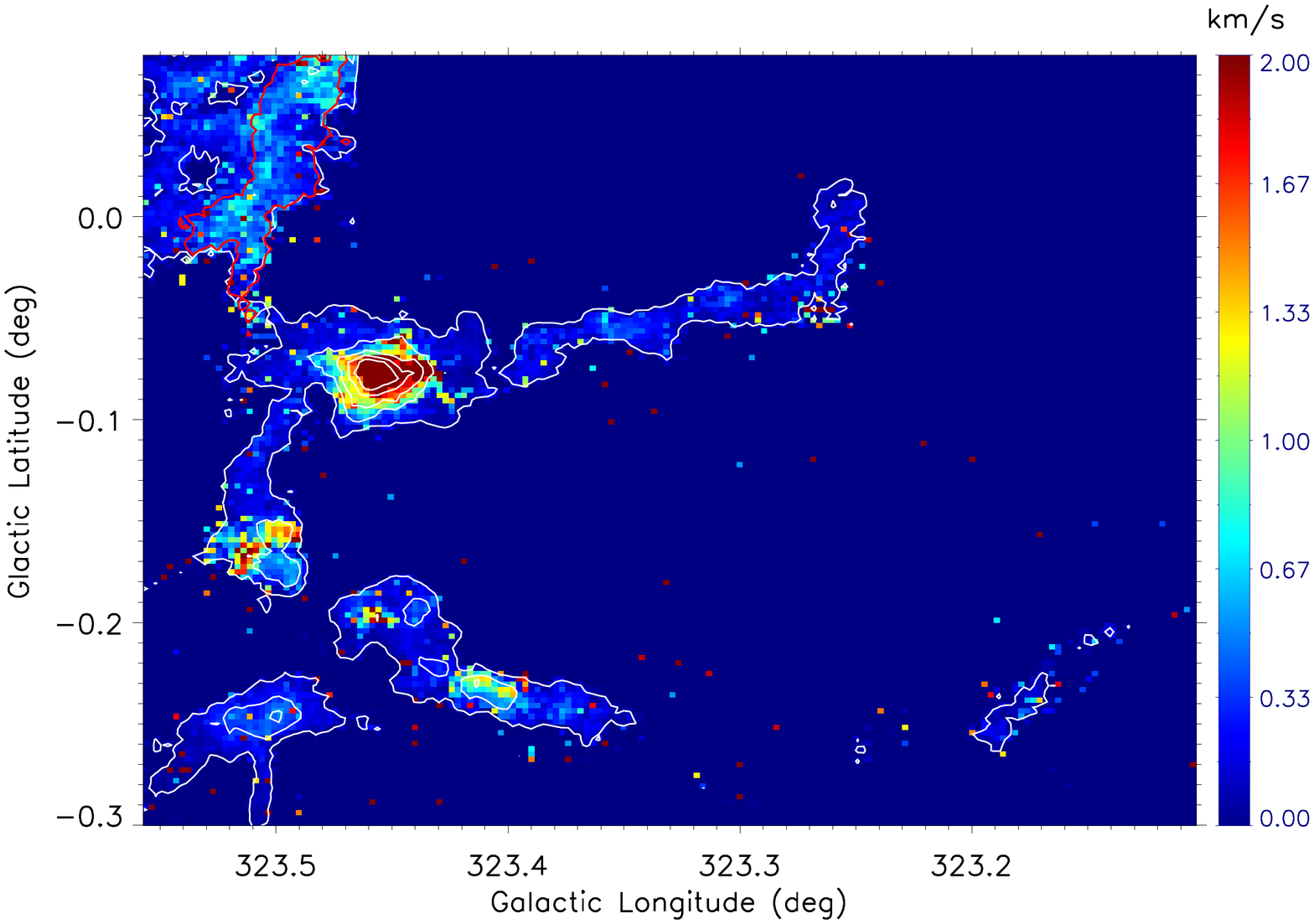}
 \caption{Moment\,2 map of G323.46-0.08. 
The colour backgrounds represents the velocity dispersion of $^{13}$CO\,($J$\,=\,2--1). 
The white contours denote the $^{13}$CO\,($J$\,=\,2--1) emission integrated velocity from -70 to -63\,km\,s$^{-1}$, which starts from a 
3$\sigma$ level of 3\,K\,km\,s$^{-1}$ and with a step of 11.5\,K\,km\,s$^{-1}$.
              }
         \label{fig2} 
  \end{figure}
%

Figure~\ref{fig1} displays the region of study centering on the high mass clump AGAL323.459-0.079 (the red cross in figure is at $l=323.459^{\circ}$, $b=-0.079^{\circ}$) \citep{Urquhart2018}. 
The left panel presents observational results of the region at different wavelengths, where red, green, and blue background trace the 24, 8, and 4.5\,$\mu$m emission, and red circles denote the H{\scriptsize II} regions identified by \citet{Anderson2014}. 
Strong 8\,$\mu$m emission is associated with the H{\scriptsize II} region. 
Class\,I (cyan crosses) and Class\,II (cyan triangles) YSOs are usually associated with the molecular cloud traced by integrated $^{13}$CO\,($J$\,=\,2--1) emission. 
The method of classification of young stars is the same as that of \citet{Gutermuth2009} and \citet{Zhou2020}.

The integrated $^{13}$CO\,($J$\,=\,2--1) emission displayed by white contours in the left panel of Figure~\ref{fig1} shows that there are three filamentary structures in the region, namely, F-NE, F-west, and F-south, all connecting to a single junction point.
They stand out more clearly in the corresponding velocity field map in the right panel. It should be noted that we derived the velocity field with the Gaussian fitting results of $^{13}$CO\,($J$\,=\,2--1) spectra for each pixel in this work. 
F-west shows a velocity gradient from its eastern end to the junction region. 
F-NE shows obvious velocity variation from the top side to the bottom side. and F-south is made of three segments, part\,1, 2, and 3. 
We can see clear velocity gradients in parts\,1, 2, and 3, although the gradient in F-south part\,3 is relatively weak. 

We noted that H{\scriptsize II} region\,2 appears between F-south part\,1 and 2, and H{\scriptsize II} region\,3 appears in the northern side of F-south part\,2. 
The northern edge of H{\scriptsize II} region 6 overlaps with F-south part\,2, and the distribution of high-velocity components well agrees with this northern edge (Figure~\ref{fig1}). 
These high-velocity components may be part of the expanding shell of H{\scriptsize II} region 6. 
The channel maps in Figure~\ref{fig2_appendix} shows that the molecular cloud associated with H{\scriptsize II} region 6 has a velocity range from -65.5 to -63\,km\,s$^{-1}$, while F-south mainly appears in the velocity range from -68 to -66\,km\,s$^{-1}$. 
It appears that F-south does not interact with the H{\scriptsize II} region 6 and they happen to coincide along the line of sight.

The fourth filamentary structure F-north with a velocity range of -70\,-\,-68.25\,km\,s$^{-1}$ appears in velocity map in Figure \ref{fig1} and seems to overlap with F-NE. 
The channel map in Figure~\ref{fig2_appendix} further confirms that this is a separate filamentary structure that seems to be associated with the central high mass clump AGAL323.459-0.079 (Figure~\ref{fig1}). 
No obvious velocity gradient is detected in F-north, possibly because F-north is perpendicular to the line of sight.

In conclusion, F-north, F-west and F-south seem to converge toward the central high-mass clump AGAL323.459-0.079 at the junction, thus constituting the hub-filament system G323.46-0.08. 
The source AGAL323.459-0.079, which was well traced by $^{13}$CO\,($J$\,=\,2--1) emission, is located at the center of this junction. 
Though another small dense clump HIGALBM323.4681-0.1011 (white cross in the right panel of Fig.\ref{fig1}) from the catalogue of \citet{Elia2017} is located south of the junction, it is not obvious in the $^{13}$CO\,($J$\,=\,2--1) map. The AGAL323.459-0.079 is the main star-forming region that is fed by the filaments and we refer to it as the "main SF" hereafter.

The systemic velocity of the hub junction is -67\,km\,s$^{-1}$ as obtained from the averaged $^{13}$CO\,($J$\,=\,2--1) spectrum, which suggests the estimated kinematic distance of the hub-filament system G323.46-0.08 to be 3.83\,$\pm$\,0.75\,kpc using the kinematic distance estimator\footnote{http://bessel.vlbi-astrometry.org/node/378} developed by \citet{Reid2019}. This new distance is smaller than that obtained by \citet{Burton2013}. Because the latest distance estimator is
based on the much improved rotation curve of the Galaxy, we use the newer distance of G323.46-0.08 in this work.

F-NE shows a clear velocity variation from its top side to its bottom side (Figure~\ref{fig1}). 
This region is the subject for a detailed study in a forthcoming paper. 
This paper will focus on the hub-filament system G323.46-0.08, namely, the region within the circled area in the right panel of Figure~\ref{fig1}. 

%
\begin{table*}
\caption{Properties of the filaments}            
\label{table:1}      
\begin{centering}                        
\begin{tabular}{l l l l l}        
\hline\hline                 
Name & sizes & Mass & M$_{line}$ & N$_{H_2}$ \\    
     & pc\,$\times$\,pc   & M$_{\odot}$ & M$_{\odot}$\,pc$^{-1}$  & 10$^{21}$cm$^{-2}$ \\
\hline                        
F-west        & 13.5\,$\times$\,0.44  & 2053\,$\pm$\,800 & 152\,$\pm$\,60    & 5.9\,$\pm$\,0.6      \\      
 F-south part\,1 & 6\,$\times$\,0.44  & 1456\,$\pm$\,600 & 242.7\,$\pm$\,100   & 6\,$\pm$\,0.6  \\
 F-south part\,2 & 9\,$\times$\,0.53  & 1879\,$\pm$\,700 & 208.8\,$\pm$\,80   & 5.1\,$\pm$\,0.5    \\
 F-south-all  & 29.4 & 3682$\,\pm$\,1500  & 125.2\,$\pm$\,50    & 5.5$\,\pm\,$0.6     \\
 F-north      & 8\,$\times$\,0.44 & 1616\,$\pm$\,600 & 202\,$\pm$\,80   &  7.4\,$\pm$\,0.7   \\
 Hub-junction & 2.4  & 3072\,$\pm$\,1200 & 439\,$\pm$\,170     &  7.5\,$\pm$\,0.8  \\
\hline                                   

\end{tabular}\\
\end{centering}
Notes: Column\,1: sub-filament name; Column\,2: deconvolved sub-filament size (length and radius, measured from $^{13}$CO\,($J$\,=\,2--1) 3\,$\sigma$\ contour). F-south is only given length due to the weak emission of F-south part\,3. The hub-junction radius 2.4\,pc is equivalent radius of a circle with the same area of junction; Column\,3: sub-filament mass; Column\,4: sub-filament mass-per-unit-length; Column\,5: averaged sub-filament column density.
\end{table*}

\subsection{Physical parameters}

Based on a modified blackbody model \citep{Kauffmann2008}, we made a spectral energy distribution (SED) fitting using the Herschel data. 
First, using the method of constrained-diffusion-decomposition \citep{Li2022}, we removed the background emission from the image data at different bands separately. 
Then we performed a spectral energy distribution fitting with the PYTHON package  HIGAL-sed-fitter\footnote{http://hi-gal-sed-fitter.readthedocs.org}. The code convolve all images at 70, 160, 250, and 350 µm to the angular resolution $36.4''$ at 500\,$\mu$m and regrid five bands images to the same pixel size $11.5''$. Finally, the SED fitting is completed based on a modified blackbody model,
      \begin{equation}
      I_\nu = B_\nu({1 - e^{-\tau_\nu}})\,,
      \end{equation}
       where the Planck function $B_\nu$ is modified by the optical depth \citep{Kauffmann2008},
\begin{equation}
\tau_\nu = \mu_{\rm H_2} m{\rm _H} K_\nu{N_{\rm H_2}} / R_{\rm gd},
\end{equation}
where $\mu_{\rm H_2}$\,=\,2.8 is the mean molecular weight, 
$m_{\rm H}$ is the mass of a hydrogen atom, 
the gas to-dust ratio $R_{\rm gd}$\,=\,100, 
and $N_{\rm H_2}$ is the H$_2$ column density. 
The dust opacity  \citep{Ossenkopf1994} is
\begin{equation}
     K_\nu = 4.0({ \nu/505\,{\rm GHz} })^\beta \,{\rm cm}^2{\rm g}^{-1},
\end{equation}
and the dust emissivity index $\beta$ was fixed to 1.75 in the fitting \citep{Wang2015}.

The derived dust temperature and $\rm H_2$ column density ($N_{\rm H_2}$) distributions of G323.46-0.08 are shown in Figure~\ref{fig3_appendix}. 
The dust temperature varies from 18 to 32\,K, and it is usually very low in filaments except for the regions where H{\scriptsize II} regions appear. 
The column density $N_{\rm H_2}$ varies from $4\times 10^{21}$ to $3.7\times 10^{22}$\,cm$^{-2}$. 
The total gas masses of filaments are calculated
using the equation
\begin{equation}
 M_{\rm H_2} = \mu m_{\rm H} \sum_{\rm i} S_{\rm i} \, N(\rm H_2)_{\rm i} \,
,\end{equation}
where S$\rm_i$ is one pixel's area, and N(\rm\,H$_2$)$_{\rm\,i}$ is obtained from SED ﬁtting. 
We get gas mass of F-north, F-west, F-south, and Hub junction is 1616, 2053, 3682, and 3072\,M$_\odot$, respectively. 
Table.\,\ref{table:1} lists the correlated parameters of sub filaments. 
Because F-north overlaps with F-NE, we cannot separate it from F-NE clearly and its true mass should be less than the value listed in Table.\,\ref{table:1}.

A total of 35 clumps from the Herschel Hi-GAL clump catalog \citep{Elia2017} and 3 clumps from the ATLASGAL dense clump catalog \citep{Urquhart2018} are located in the hub filament G323.46-0.08 (see left panel of Figure~\ref{fig3_appendix}). 
These clumps represent sites of ongoing star formation and contain pre-stellar and proto-stellar cores.
The sources HIGALBM323.4594-0.0789 and AGAL323.459-00.079, as well as HIGALBM323.4927-0.1781 and AGAL323.494-00.179, and HIGALBM323.5076+0.0455 and AGAL323.506+00.046 are identical sources. 
All 35 clumps and their parameters were recalculated using the new distance are listed in Table~2. 
The mass of the central high-mass clump AGAL323.459-0.079 (main SF) from \citet{Elia2017} is much lower than the result from \citet{Urquhart2018}, because the former used a smaller radius (see Table.\ref{table:2}). We recalculated the mass with the same radius of \citet{Urquhart2018}  
and obtained a mass of 1100\,M$_\odot$, which is comparable to  1170\,M$_\odot$ obtained by \citet{Urquhart2018}.

\subsection{The velocity dispersion}

Figure~\ref{fig2} displays the distribution of the $^{13}$CO\,($J$\,=\,2--1) velocity dispersion of the region. 
The main SF has the largest velocity dispersion of 3.2\,km\,s$^{-1}$, which may be attributed to ongoing star formation activity but also to a possible outflow in the source \citep{yang2022}. 
The source F-west has a very small and uniform velocity dispersion, even at the regions where H{\scriptsize II} regions 1 and 4 appear. 
On the other hand, F-south also has a similar small velocity dispersion, but it is large in the H{\scriptsize II} regions 2, 3, and 6. 
The source F-north has a relative large velocity dispersion at its eastern edge, which may be attributed to the gas mixing of F-north and F-NE or interaction between them.

The contributions of thermal and non-thermal motions to the velocity dispersion may be determined following \citet{Myers1983} and \citet{Tang2017, Tang20182,Tang2018,Tang2021}:
\begin{equation}
\sigma_{NT}^2 = \sigma_{obs}^2 - \sigma_T^2 
,\end{equation}
and
\begin{equation}
\sigma_{NT} = \sqrt{\cfrac{\Delta\,v_{obs}^2}{8\,\ln{2}}-\cfrac{k_B\,T_{kin}}{m_{obs}}},  
\end{equation}
\newline where $\sigma_{NT}$, $\sigma_{obs}$, and $\sigma_T$ are the non-thermal, observed, and thermal velocity dispersions, respectively. 
In addition, $\Delta\,v_{obs}$ represents the observed full width at half maximum (FWHM) line width of $^{13}$CO\,($J$\,=\,2--1), $k_B$ is the Boltzmann constant, $T_{kin}$ is the kinematic temperature of the gas, and $m_{obs}$ is the mass of the observed molecule for $^{13}$CO\,($J$\,=\,2--1). 
Assuming the kinematic temperature of gas is equal to the dust temperature (Figure~\ref{fig3_appendix}), the average thermal dispersion is about 0.076, 0.079, and 0.083\,km\,s$^{-1}$ for F-north, F-west, and F-south, respectively. 
The corresponding non-thermal velocity dispersion are about 0.46\,(F-north), 0.25\,(F-west) and 0.53\,km\,s$^{-1}$\,(F-south), which are greater than the thermal dispersion. 
Therefore, the velocity dispersion of these filaments is dominated by non-thermal motions.

    \begin{figure}
  \centering
  \includegraphics[trim={0cm 0cm 0cm 0cm},clip,width=9cm]{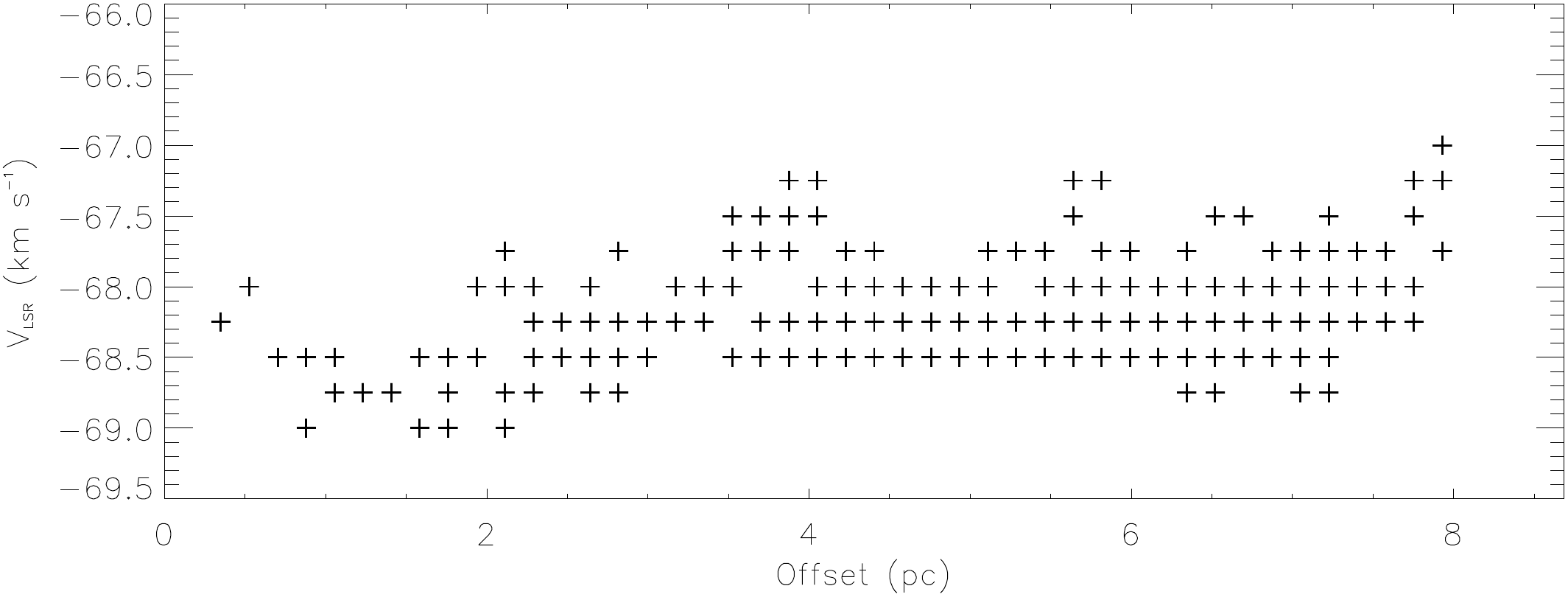}
      \caption{ $^{13}$CO\,($J$\,=\,2--1) line velocity along F-north. The x-axis is taken along the red dotted line in the right panel of Figure~\ref{fig1}, with the zero position starting from its intersection at the junction.}
         \label{fig3_2}
  \end{figure}
  
  \begin{figure*}
  \centering
  \includegraphics[trim={0.8cm 1cm 0cm 1.2cm},clip,width=19cm]{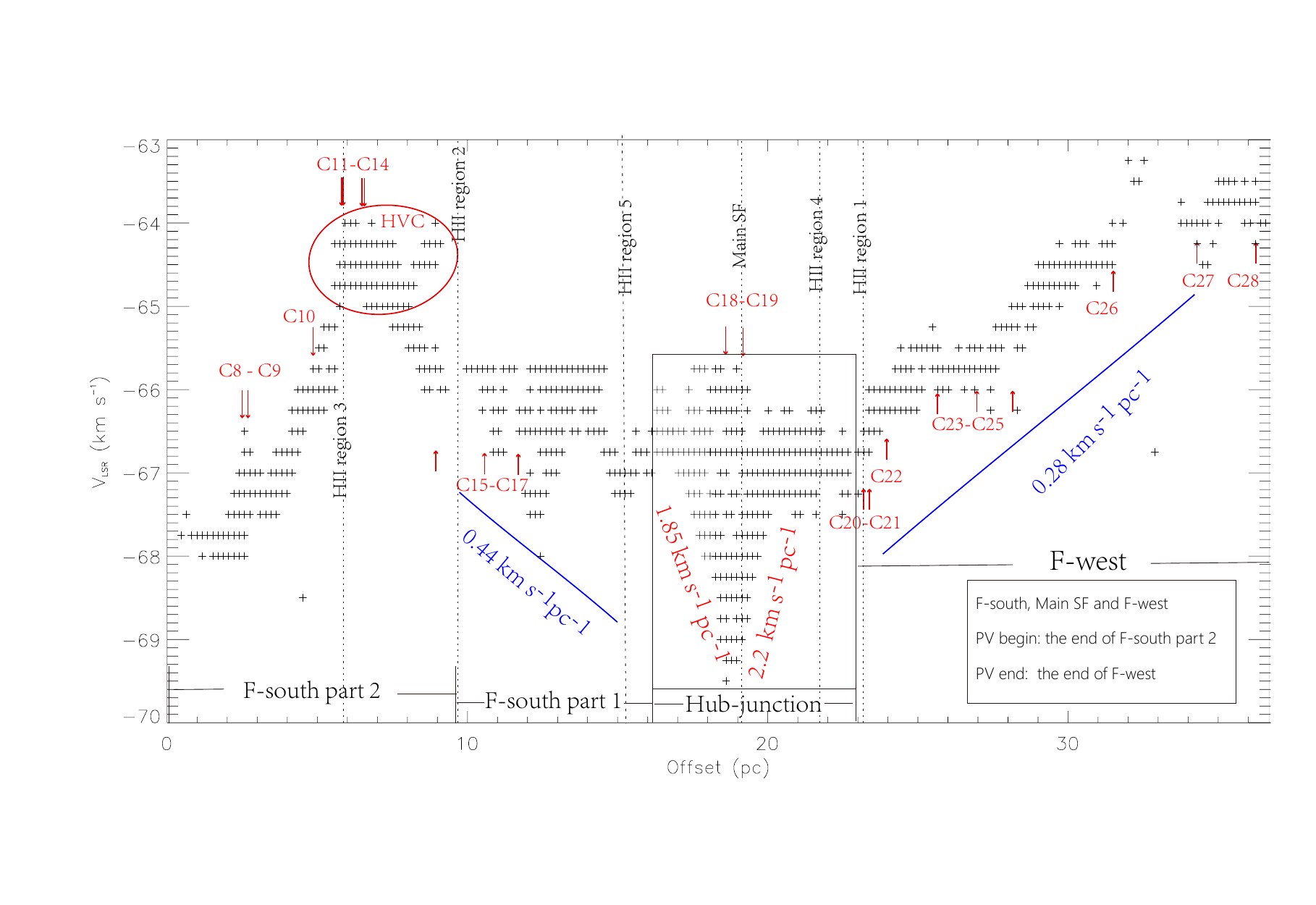}
      \caption{ $^{13}$CO\,($J$\,=\,2--1) line velocity along F-south to F-west. 
The blue lines indicate the velocity gradient in the F-west and F-south part\,1. 
The vertical black dotted lines indicate the position of H{\scriptsize II} region along the filament, and red arrows indicate the location of the numbered pre-stellar and proto-stellar clumps located along the filament. The data points circled by the red ellipse is the HVC high-velocity component defined in the right panel of Figure~\ref{fig1}.
              }
         \label{fig3}
  \end{figure*}

  \begin{figure}
  \centering
  \includegraphics[trim={1.2cm 0.5cm 0.3cm 0cm},clip,width=11cm]{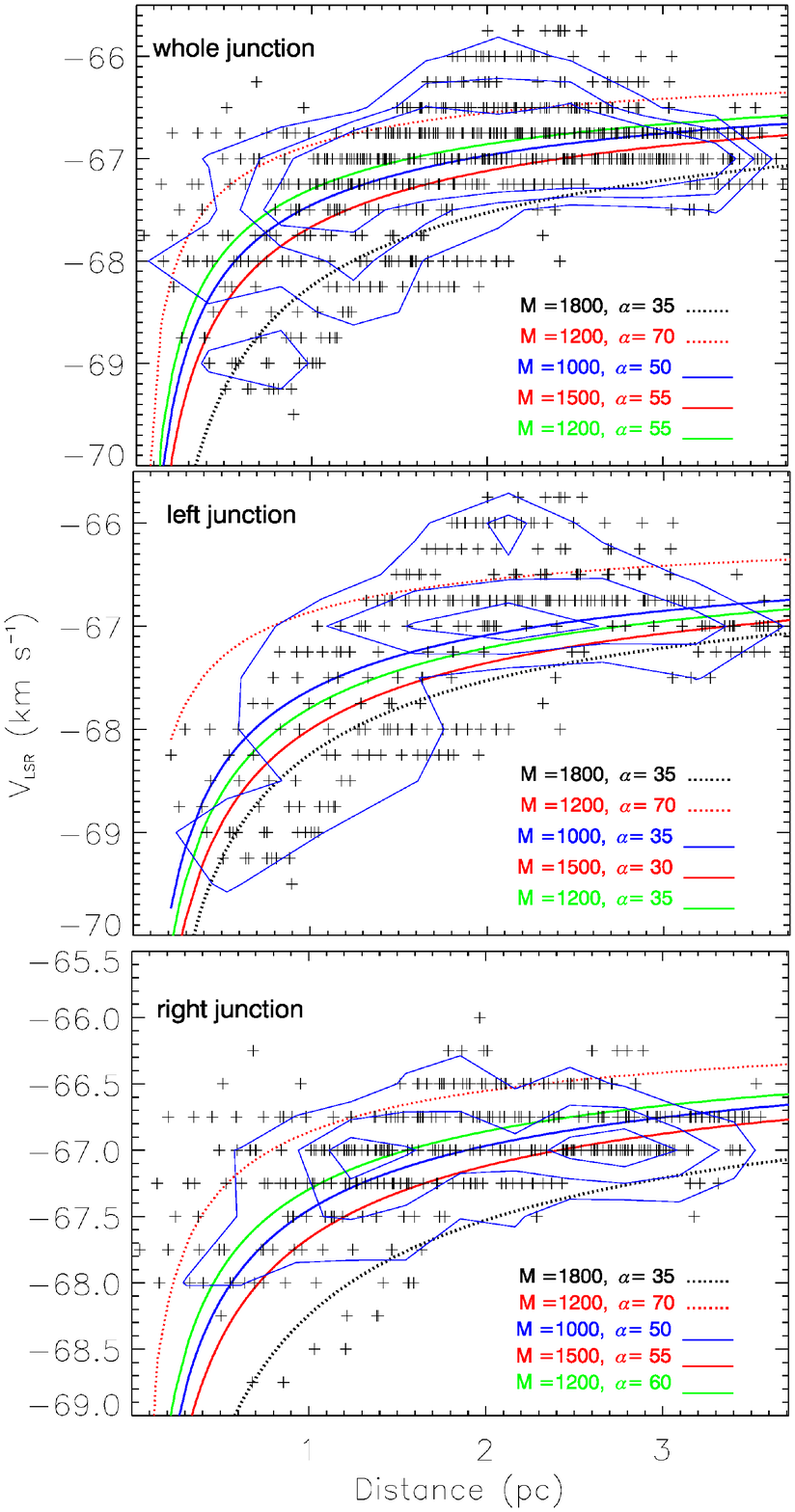}
 \caption{Gas velocity structure as a function of the distance to the center clump ($l = 323.459^{\circ}$, $b = -0.079^{\circ}$). 
The data points from the top to the bottom panels of the diagram are from the whole junction, the left half of the junction ($l > 323.46^{\circ}$) and the right half of the junction ($l < 323.46^{\circ}$), respectively. 
For all panels different fitted lines describe the expected velocity profile for a free-falling particle in a series of potential wells with different masses $M$ observed at different angles $\alpha$ following Eq.\,(4). 
The parameters are showed at the lower right corner of each panel, the masses are in M$_\odot$ and angles are in degrees. The blue contours are the density-plot of the cross points.}
         \label{fig4} 
  \end{figure}
%

\section{Discussion}

\subsection{Global velocity gradient }

To clearly show the kinematics along the filament and the relation between F-north, F-west, F-south, and the main SF, two position-velocity diagrams have been created using the $V_{LSR}$ at all pixels with $S/N>5$ in $^{13}$CO\,($J$\,=\,2--1) emission as a function of their corresponding offset along the filaments (Figures~\ref{fig3_2} and~\ref{fig3}). 
For F-north in Figure~\ref{fig3_2}, the x-axis is taken along the red dotted line in the right panel of Figure~\ref{fig1}, with the zero position starting from its intersection at the junction. 
For F-south and F-west in Figure~\ref{fig3}, the x-axis starts at the end of F-south part\,2 and along the white interrupted dashed line across the main SF to the end of F-west. These lines are determined by connecting the peak intensity points of $^{13}$CO\,($J$\,=\,2--1) along the filaments. 
In Figure~\ref{fig3}, F-west shows a continuous velocity gradient from its end to the main SF. 
By least-squares fitting, we obtained a large scale velocity gradient of 0.28\,$\pm$\,0.06\,km\,s$^{-1}$\,pc$^{-1}$ along F-west. 

The velocity variation along F-south is more complex. 
F-south part\,1 as a whole shows a velocity gradient approaching the junction region, while F-south part\,2 shows an opposite velocity variation that may result from different projection angles. The northern part of the H{\scriptsize II} region\,6 seems to overlap with F-south part\,2 along the line of sight, but these components cannot be separated clearly. 
The high-velocity components in the red circle marked as HVC in the bottom panel of Figure~\ref{fig3} appear as an arc structure and may include part of H{\scriptsize II} region\,6's expanding shell (also see  Figure~\ref{fig1}). 

The velocity gradient along F-south part\,1 is 0.44\,$\pm$\,0.1\,km\,s$^{-1}$\,pc$^{-1}$, which obtained by least-squares fitting. 
The velocity gradients detected in F-west and F-south part\,1 are comparable to those detected in other hub-filament systems, for instance, 0.8\,$\pm$\,0.1\,km\,s$^{-1}$\,pc$^{-1}$ in the DR 21 South Filament \citep{hu2021} and 0.15--0.6\,km\,s$^{-1}$\,pc$^{-1}$ in SDC13 \citep{Peretto2014}. 
Above results suggest that F-west and F-south may be channeling materials into the main SF at the junction region. 
Furthermore, a characteristic V-shaped structure (see hub junction in Figure~\ref{fig3}) appears around the main SF. 
This feature is in good agreement with the simulation results of \citet{kuznetsova2018} and \citet{gomez2014}, and is similar to what is detected in Orion by \citet{hacar2017}. 
The V-shaped structure indicates that accelerated collapse is taking place around the main SF.  

The velocity gradients of the two sides of the V-shaped structure are 1.85 and 2.2\,km\,s$^{-1}$\,pc$^{-1}$, respectively (Figure~\ref{fig3}). 
They are much less than the simulation result 4--7\,km\,s$^{-1}$\,pc$^{-1}$ of \citet{kuznetsova2018} and the observation result 5--7\,km\,s$^{-1}$\,pc$^{-1}$ in Orion \citep{hacar2017}. 
Considering that $^{13}$CO\,($J$\,=\,2--1) is optically thick in the dense regions at the core, the infall motion of dense gas with a large velocity gradient cannot be detected and only the outer parts of the clump with a small gradient that are detected as gravitationally collapsing. 
It is evident more observations with dense molecular tracers and high angular resolution are needed for further validation. In addition, the projection effect may also make the velocity gradient relatively low. 

 F-north has no obvious velocity gradient (Figure~\ref{fig3_2}), perhaps because it is close to perpendicular to the direction of the line of sight.  
 However, there are velocity fluctuations along F-north.

\subsection{Gravitational collapse of the junction region}

Accelerated collapse is taking place close to the main SF as seen in Figure~\ref{fig3}. 
If this characteristic feature is caused by gravity, we can use a simple model to plot the gas velocity structure as a function of the distance. All points in the junction with $^{13}$CO\,($J$\,=\,2--1) emission greater than 5\,$\sigma$ were selected. Assuming the position of high mass clump AGAL323.459-0.079, that is, $l_0=323.459^{\circ}$ and $b_0=-0.079^{\circ}$, is the offset (0,0), the gas velocity structure of the whole junction as a function of distance to the offset (0,0) is plotted in the top panel of Figure~\ref{fig4}. 
The distance R of the x-axis is obtained through $R = \sqrt{(l-l_0)^2 + (b-b_0)^2}$. 
There is a clear velocity gradient in Figure~\ref{fig4}. 

Assuming the model of a point-like mass in free-fall, the observed velocity $V_{LSR}$ at a given impact parameter $p$ can be described by the following relation \citep{hacar2017}:
      \begin{equation}
V_{LSR}(p) = V_{sys,0} + V_{infall}(p)\,\cdot\,cos(\alpha)
      ,\end{equation}
\noindent where V$_{sys,0}$ is the initial system velocity, and
\begin{equation}
V_{infall}(p) = -\sqrt{2GM/R} = -\sqrt{2GM/(p/sin\alpha)}
,\end{equation}
\noindent  is the infall velocity in a potential of a mass $M$ as a function of the distance $R$ to its center with $\alpha$ as  the orientation angle between the direction of infall motion and the direction of the line of sight. 
In Figure~\ref{fig4} the expected velocity profiles are displayed for particles in free-fall in different potential wells and with different orientation angles. 
The model results show masses between 1000-1500\,M$_\odot$ and an orientation angle of about 55$^{\circ}$ are well consistent with observations, which are showed as color solid lines in Figure~\ref{fig4}. 
This agrees with the fact that the main SF has a mass of 1100\,M$_\odot$. The result shows that the gradient is very small for the points with distances larger than $\sim$2 pc, but increases rapidly as the points approach the center of the main SF. Namely, motions of the points from the main SF or the region close to it is dominated by the gravity, and is in gravitational collapse. The corresponding density plot (blue contours in Figure~\ref{fig4}) also shows the same trend. This is consistent with the theoretical results that filamentary collapse is a lot slower than spherical collapse \citep{pon2012,clarke2017}. 

Because F-west and F-south transfer materials into the junction region from two directions, the materials from both sides of the junction may fall into the main SF at different orientation angles respectively. 
To test this idea, we used the same method to plot the velocity structure for the left and right half of the junction separately. 
The results are shown in middle and bottom panel of Fig.\ref{fig4}.
Indeed, the orientation angles $\alpha$ for the left and right half of junction are different, specifically, they are about 35$^\circ$ and 55$^\circ$, respectively. 
However, model fitting for the left and right half of junction both obtain a mass of 1000-1500\,M$_\odot$ for the main SF. Model fitting for the whole junction or for the two sides both present a systemic velocity $V_{sys,0}$ = -65.8\,km\,s$^{-1}$,
which is similar to the observed gas velocities at the beginning of gas acceleration zones in the V-shaped structure (see Figure~\ref{fig3}). 
Above results indicate that although F-west and F-south part\,1 may transfer materials into the junction at different angles, they are both attracted by the same gravitational source at the junction. 
As pointed out by \citet{hacar2017}, the above results suggest that the accelerated motions identified toward the main SF could actually correspond to the kinematic signature generated by its gravitational collapse.

\subsection{Accretion from the filament }
Following \citet{Kirk2013}, the accretion rate from the filament could be estimated from the velocity gradient using a simple cylindrical model, $\Dot{M} = (\nabla\,V\,M)/\tan(\beta)$. Here, $M$ is the cylinder's mass and $\beta$ is an inclination angle of the cylinder relative to the plane of the sky and assumed to be 45$^{\circ}$. 
The accretion rates of F-west, and F-south part\,1 are 575 and 641 M$_\odot$\,Myr$^{-1}$, respectively. 
Since F-north has no velocity gradient, its accretion rate cannot be calculated in this way. 
Therefore, at least $\sim$ 1216\,M$_\odot$ of gas will be channeled into the junction region of hub-filament G323.46-0.08 in 1\,Myr if the current accretion rates are sustained. Such a accretion rate is higher than that of \citet{Kirk2013} 30\,M$_\odot$ and \citet{Yuan2018} 440\,M$_\odot$, and is comparable to 1100M$_\odot$\,Myr$^{-1}$ in DR21SF \citep{hu2021}. Therefore, filamentary accretion flows may play an important role for supplying the material to spur high-mass clump and star formation therein. 

\subsection{Other possibilities}
Except the inflows along the filament, the rotation and outflow may also lead to a radial velocity gradient along the filamentary molecular cloud. 
To take into account the effect of rotation, F-west may be taken as a homogeneous rigidly rotating cylinder with a total rotation energy to be calculated using the following equation \citep{Kirk2013,wu2018}:
      \begin{equation}
E_{rot} = \frac{1}{6}\,(M\,L^{2})\,(\nabla\,V)^{2}, 
      \end{equation} 
\noindent where the second term is the moment of inertia for a cylinder rotating longways along its axis, and the final term represents the angular velocity squared.
The {$E_{rot}$ obtained for F-west is about 4889\,$M{_\odot}$\, km$^{2}$\,s$^{-2}$.
The gravitational binding energy is calculated by the equation:
      \begin{equation}
E_{grav} = G\,M\,M_{clust}/L,
      \end{equation}
\noindent where $M_{clust}$ is the mass of whole junction of 3072\,M${_\odot}$. 
The gravitational binding energy, $E_{grav}\,\simeq$ 2077\,M${_\odot}$\, km$^{2}$\,s$^{-2}$, is less than the rotation energy. Therefore, we cannot rule out the possibility that F-west may be rotating. However, even if F-west is rotating around the junction, the rotation energy of F-west close to the junction will be much smaller than the gravitational binding energy. F-west would be dominated by the gravity quickly as the matter approaches the junction. 
  
We checked the distributions of red and blue line wings of $^{13}$CO\,($J$\,=\,2--1), and proved that F-west could not be a red- or blue-shifted lobe of the outflow detected in main SF. We thus suggest the velocity gradient in F-west indicates an accretion ﬂow along the
ﬁlament, which is induced by the gravitational field of the entire junction region of the hub-filament system G323.46-0.08. 
This is also true for F-south part\,1.

\subsection{Star formation}

Assuming cylindrical hydrostatic equilibrium, \citet{andre2014} presented the thermal critical mass
per unit length for a gas filament as:
\begin{equation}
 M_{line,crit} = 2\,c_s^2/G = 16.7\,(T/10\,K)\,M_\odot\,pc^{-2}. 
 \end{equation}
 In this manner, the values for $M_{line,crit}$ for F-north, F-west, and F-south are 31.7, 34.4, and 36.7\,$M_\odot$\,pc$^{-1}$ for  dust temperature 19.6, 20.6, and 22\,K, respectively. 
The mass per unit length for the three filaments ranges from 125.2 to 242.7 M$_\odot$\,pc$^{-1}$ (see Table.\,\ref{table:1}). 
These filaments are supercritical with $M_{line}$ significantly larger than the critical value. 

Thermally, supercritical filaments are usually associated with dense clumps and star formation activities \citep{andre2014}. At least 35 dense clumps from the Hi-GAL dense clump catalog are located along the filaments (see left panel of Figure~\ref{fig3_appendix} and  here we only consider clumps located in hub filament system G323.46-0.08) and most of these are pre-stellar clumps. 
Pre-stellar clumps are usually located in the filaments, while proto-stellar clumps often appear near H{\scriptsize II} regions (see left panel of Figure~\ref{fig3_appendix}). 
The Class\,I (cyan crosses) and II YSOs (cyan triangles) sources in Figure~\ref{fig1}(left) are also distributed on or nearby the filaments, which suggests that their formation is also correlated with the filaments.

There are 20 clumps numbered from C8 to C28 (see table 2) in Figure~\ref{fig3} and 5 H{\scriptsize II} regions numbered from 1 to 5 located in F-west and F-south as indicated in Figure~\ref{fig1}. 
These co-locations are consistent with the global hierarchical collapse model, which predicts that filaments transfer materials into the central hub and form high mass stars or star clusters \citep{vazquez2017}. 
Simultaneously, the filaments themselves will also fragment into dense clumps or cores and form low- and intermediate-mass stars locally, which results in local accretion flows along the fragmented filaments to those dense clumps. 
As \citet{Hacar2011} showed in their figure 12, the PV diagram along the filament may show obvious velocity undulation around dense clumps, which trace the red- or blue-shifted accretion flows toward the dense clumps. 
The PV diagram along the filaments F-south and F-west ( Fig.\ref{fig3}) shows clear velocity undulation. 
The seesaw velocity patterns around clumps C\,8\,-C\,9, C\,10, H{\scriptsize II}\,2, C\,20\,-\,C\,21, and C\,23\,-\,C\,25  suggest that they are also accreting materials from the filament. 
However, velocities near some clumps do not show seesaw patterns due to projection effects and the complex structure of the filament itself. 
On the other hand, our data has a limited resolution and sensitivity, and most clumps were not clearly detected in $^{13}$CO\,($J$\,=\,2--1) emission and their kinematics. 
The seesaw pattern of the velocity structure on a large scale should be the result of gas flows along the filament toward the clumps or cores \citep{Hacar2011}. 

These results support the global hierarchical collapse model \citep{vazquez2017}, where materials converge through filaments towards a gravitational potential center, while simultaneously forming stars locally on the way to the convergence center. 

\section{Conclusions}

Using the gas kinematics derived from $^{13}$CO\,($J$\,=\,2--1) spectral lines, we found a giant hub-filament system G323.46-0.08. Our results are given below.
 \begin{enumerate}
\item The G323.46-0.08 consists of three filaments named F-north, F-west, and F-south which have lengths of 8, 13.5, and 29.4\,pc and masses 1616, 2053, and 3682\,M$_\odot$, respectively. Here, 
N$_{H_2}$ varies from 4\,$\times$\,10$^{21}$ in the filaments to 3.7\,$\times$\,10$^{22}$\,cm$^{-2}$ in the main SF. 
All three filaments are supercritical with M$_{line}$ significantly larger than the critical value $M_{line,crit}$ indicating that the filaments are unstable. 
There are YSOs and 35 clumps with pre-stellar and proto-stellar cores lying mainly along the filaments of G323.46-0.08. 
The seesaw patterns near most dense clumps in the PV diagram suggests that mass accretion also occurs along the filament toward the clumps. 
The large velocity dispersion and a possible outflow in the main SF indicate that high-mass star formation appears to be ongoing. 
 
 \item Very clear velocity gradients of 0.28 and 0.44\,km\,s$^{-1}$\,pc$^{-1}$ are found along for F-west and F-south part\,1, respectively, which suggests the channeling of materials towards the junction of hub-filament system G323.46-0.08. 
Furthermore, the characteristic V shape structure detected in the PV diagram at the hub-junction is in good agreement with a gravitational collapse model. 
Our best-fitting parameters with a hub-junction mass between 1000 and 1500\,\,M$_\odot$ are consistent with the observed mass 1100\,\,M$_\odot$ for AGAL323.459-0.079. 
 
\item An estimate of the minimum accretion rate onto G323.46-0.08 is 1216\,M$_\odot$\,Myr$^{-1}$, obtained by channeling gas from filaments feeding onto the central clump.  
Filamentary accretion flows therefore appear to be an important mechanism for feeding high-mass star formation in the central high-mass clump AGAL323.459-0.079 and for intermediate star formation in clumps along the filaments.

 \end{enumerate}

In a brief, G323.46-0.08 is accreting materials through the filaments onto AGAL323.459-0.079. AGAL323.459-0.079 is under gravitational collapsing and forming high-mass stars. Many dense clumps in the filaments also show evidences of forming stars.

\begin{acknowledgements}
This work was mainly supported by the National Key R\&D Program of China under grant No.2022YFA1603103 and the National Natural Science foundation of China (NSFC) under grant No.11973076. It was also partially supported by the NSFC under grant Nos. 12173075, and 12103082, the Natural Science Foundation of Xinjiang Uygur Autonomous Region under grant Nos. 2022D01E06 and 2022D01A359, the Tianshan Talent Program of Xinjiang Uygur Autonomous Region under grant No. 2022TSYCLJ0005, the Chinese Academy of Sciences (CAS) “Light of West China” Program under Grant Nos. 2020-XBQNXZ-017, 2021-XBQNXZ-028, 2022-XBJCTD-003 and  xbzg-zdsys-202212, 
the Regional Collaborative Innovation Project of Xinjiang Uyghur Autonomous Region under grant No.2022E01050, and the Science Committee of the Ministry of Science and Higher Education of the Republic of Kazakhstan grant No. AP13067768.   
W.B. has been supported by the Chinese Academy of Sciences President International Fellowship Initiative by Grant No. 2022VMA0019 and 2023VMA0030.

\end{acknowledgements}

.

\bibliography{ref}

\begin{appendix} 
\section{$^{13}$CO\,($J$\,=\,2--1) velocity channel maps}
  \begin{figure*}
  \centering
  \includegraphics[trim={0.5cm 0cm 1cm 1cm},clip,width=20cm]{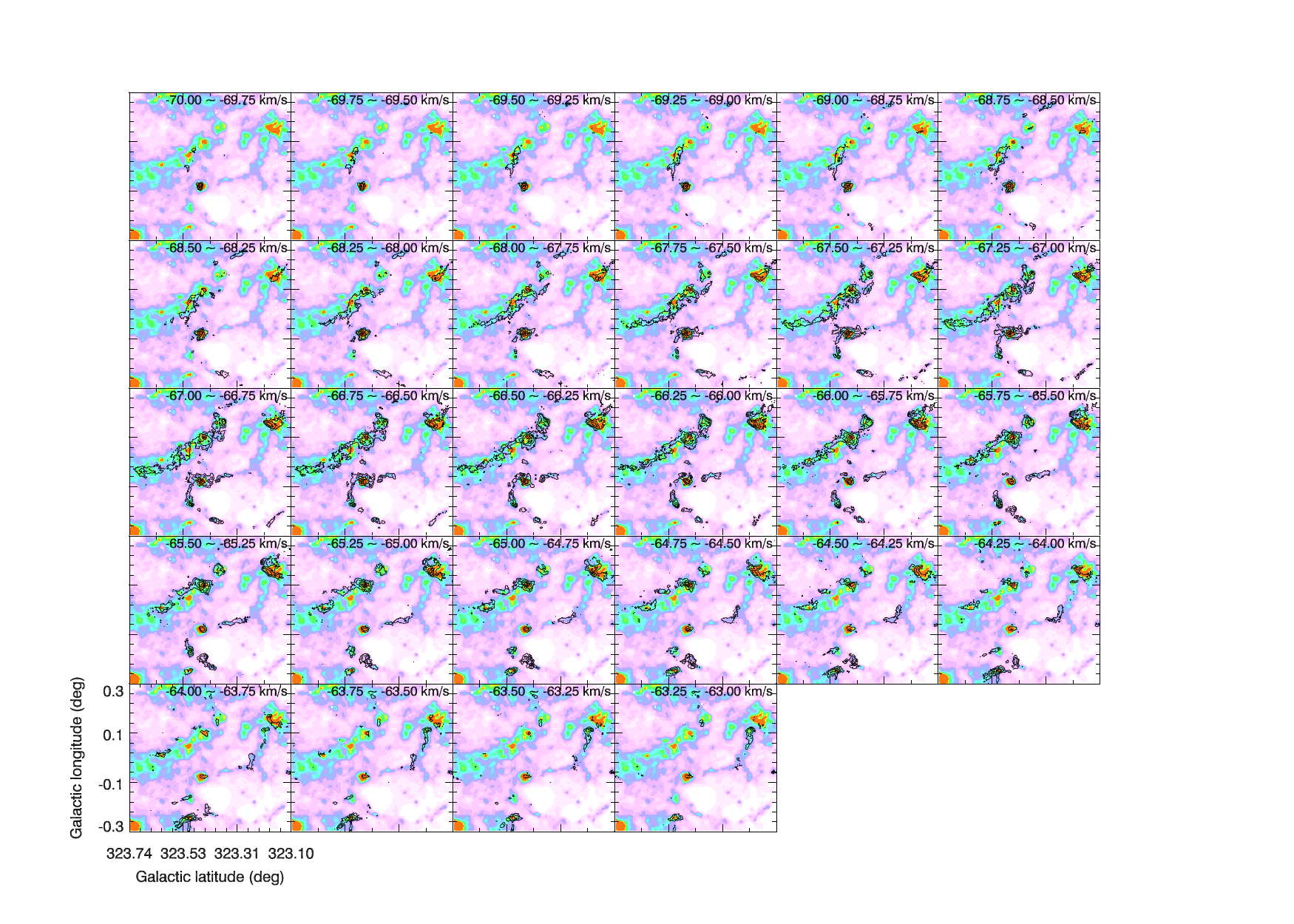}
  \caption{Channel maps of the $^{13}$CO\,($J$\,=\,2--1) emission for G323.46-0.08 with black contour levels at $3$, $6$, $9$, $12$, and $15$ and 18\,K\,km\,s$^{-1}$.
  The background is the H$_2$ column density distribution.} 
         \label{fig2_appendix} 
  \end{figure*}

\section{Table~\ref{table:2}}
\begin{table*}
\caption{Parameters of dust clumps associated with G323.46-0.08.}         
\label{table:2}      
\begin{centering}                         
\begin{tabular}{p{5.8cm}p{0.8cm}p{1.0cm}p{1.8cm}p{1.3cm}p{1.3cm}p{2.6cm}}
\hline\hline                 
Name & Diam & Mass & Evol & Diam(new) & Mass(new) &surface density  \\    
     & pc   & M$_\odot$ & &pc & M$_\odot$ & g\,cm$^{-2}$\\
\hline                        
HIGALBM323.1392-0.2052 (ID:C1) & 1.282 &  1205.02 & prestellar  & 0.433 & 137.44  & 0.195\\
HIGALBM323.1537-0.2118 (ID:C2) & 1.900 &  1254.33 & prestellar  & 0.642 & 143.08  & 0.092 \\
HIGALBM323.1746-0.2310 (ID:C4) & 0.839 &  782.20  & prestellar  & 0.282 &  87.96  & 0.295  \\
HIGALBM323.2436-0.2680 (ID:C7) & 0.973 &  630.51  & protostellar& 0.374 &  92.85  & 0.177   \\
HIGALBM323.2622-0.0213 (ID:C27)& 1.177 &  745.49  & prestellar  & 0.476 &  122.06 & 0.143   \\
HIGALBM323.2647+0.0094 (ID:C28) & 1.165 &  252.27  & prestellar  & 0.471 &   41.30 & 0.049   \\
HIGALBM323.3619-0.0512 (ID:C23) & 1.376 & 1534.50  & prestellar  & 0.453 &  165.97 & 0.216    \\
HIGALBM323.3771-0.2326 (ID:C8) & 1.118 &   356.56 & prestellar  & 0.456 &   59.36 & 0.076 \\
HIGALBM323.3872-0.0608 (ID:C22) & 1.188 &  1186.24 & prestellar  & 0.492 &  203.75 & 0.224  \\
HIGALBM323.3976-0.0697 (ID:C21) & 1.584 &   953.49 & protostellar& 0.657 &  163.77 & 0.101 \\
HIGALBM323.3985-0.0652 (ID:C20) & 1.491 &   532.55 & protostellar& 0.618 &   91.47 & 0.064  \\
HIGALBM323.4117-0.2313 (ID:C10) & 1.321 &  3937.63 & protostellar& 0.539 &  655.48 & 0.600 \\
HIGALBM323.4273-0.2328 (ID:C11) & 0.946 &   129.67 & protostellar& 0.386 &  21.59  & 0.039 \\
HIGALBM323.4360-0.2267 (ID:C13) & 1.800 &  1769.43 & protostellar& 0.729 &  290.57 & 0.145 \\
HIGALBM323.4376-0.1960 (ID:C14) & 0.782 &   255.97 & protostellar& 0.316 &  42.04  & 0.111\\
HIGALBM323.4594-0.0789 (ID:C19) & 0.318 &  1068.83 & protostellar& 0.277 & 815.27  & 2.806 \\
(AGAL323.459-00.079)   & 1.784 &  1276.44 & massive star formation       & 1.708 & 1170.25 & 0.113  \\
HIGALBM323.4681-0.1011 (ID:C18) & 1.266 &  2436.76 & protostellar& 0.528 & 424.02  & 0.404 \\
HIGALBM323.4744-0.1780 (ID:C15) & 0.725 &  43.55   & protostellar& 0.294 &   7.15  & 0.022\\
HIGALBM323.4907+0.0234 & 1.067 &  1977.23 & prestellar  & 0.442 & 338.29  & 0.462 \\
HIGALBM323.4926+0.0736 & 0.492 &  1659.51 & prestellar  & 0.429 & 1257.74 & 1.827 \\
HIGALBM323.4927-0.1781 (ID:C16) & 1.488  & 825.51 & protostellar & 0.603  & 135.57 & 0.099 \\
(AGAL323.494-00.179)   & 0.380 &    88.31 & YSO         & 0.393 &   94.43 & 0.163  \\
HIGALBM323.4953-0.1603 (ID:C17) & 1.588 &  4888.24 & protostellar& 0.662 &  850.61 & 0.516  \\
HIGALBM323.5076+0.0455 & 0.632  & 951.40& prestellar & 0.561 & 746.95& 0.633   \\
(AGAL323.506+00.046)   & 0.426 &   645.65 & protostellar& 0.441 &  690.44 & 0.951   \\
HIGALBM323.5179+0.0023 & 0.533 &   101.71 & protostellar& 0.221 &   17.64 & 0.095  \\

HIGALBM323.1593-0.2265 (ID:C3) & 0.977 &  1851.75 & prestellar  & 0.330 &  211.22 & 0.516  \\
HIGALBM323.1844-0.2427 (ID:C5)& 1.126 &   400.77 & prestellar  & 0.378 &   45.06 & 0.084  \\
HIGALBM323.1885-0.2574 (ID:C6)& 1.656 &   567.08 & prestellar  & 0.555 &   63.76 & 0.055  \\
HIGALBM323.2754-0.0409 (ID:C26) & 0.898 &   201.29 & prestellar  & 0.370 &   34.34 & 0.066  \\
HIGALBM323.3238-0.0491 (ID:C25) & 1.114 &   960.17 & prestellar  & 0.366 &   103.84 & 0.206  \\
HIGALBM323.3426-0.0569 (ID:C24) & 1.355 &   531.09 & prestellar  & 0.446 &   57.45 & 0.077  \\
HIGALBM323.3790-0.2458 (ID:C9)& 1.812 &  1001.69 & prestellar  & 0.752 &  172.31 & 0.081  \\
HIGALBM323.4271-0.2050 (ID:C12) & 0.879 &    55.34 &   starless  & 0.359 &   09.21 & 0.019  \\

HIGALBM323.5023-0.0470 & 1.107 &   248.95 & prestellar  & 0.461 &   43.16 & 0.054  \\
HIGALBM323.5117-0.0034 & 1.247 & 10550.60 & prestellar  & 0.519 & 1829.16 & 1.805  \\
HIGALBM323.5117-0.0612 & 0.721 &   292.51 & prestellar  & 0.300 &   50.70 & 0.150  \\
\hline                                   
\end{tabular}
\end{centering}
Note. The first, second, and third columns list the names, diameters, and masses of the clumps in the Herschel Hi-Gal clump catalog \citep{Elia2017}. 
 The fourth columns indicate the evolutionary stage of the clumps.
   The fifth and sixth columns indicate the new diameters and masses calculated for a distance of 3.83\,kpc. 
   The seventh column presents the clump surface density. 
   AGAL323.459-00.079, AGAL323.494-00.179, and AGAL323.506+00.046 are the clumps that fall into the G323.18+0.15 in the ATLASGAL dense clump catalog. 
   The sources AGAL323.459-00.079 and HIGALBM323.4594-0.0789, AGAL323.494-00.179  and HIGALBM323.4927-0.1781, HIGALBM323.5076+0.0455 and AGAL323.506+00.046 are duplicated.
\end{table*} 

\section{A selection of typical $^{13}$CO\,($J$\,=\,2--1) spectra of G323.46-0.08}

    \begin{figure*}
  \centering
  \includegraphics[trim={0cm 1cm 2cm 2cm},clip,width=20cm]{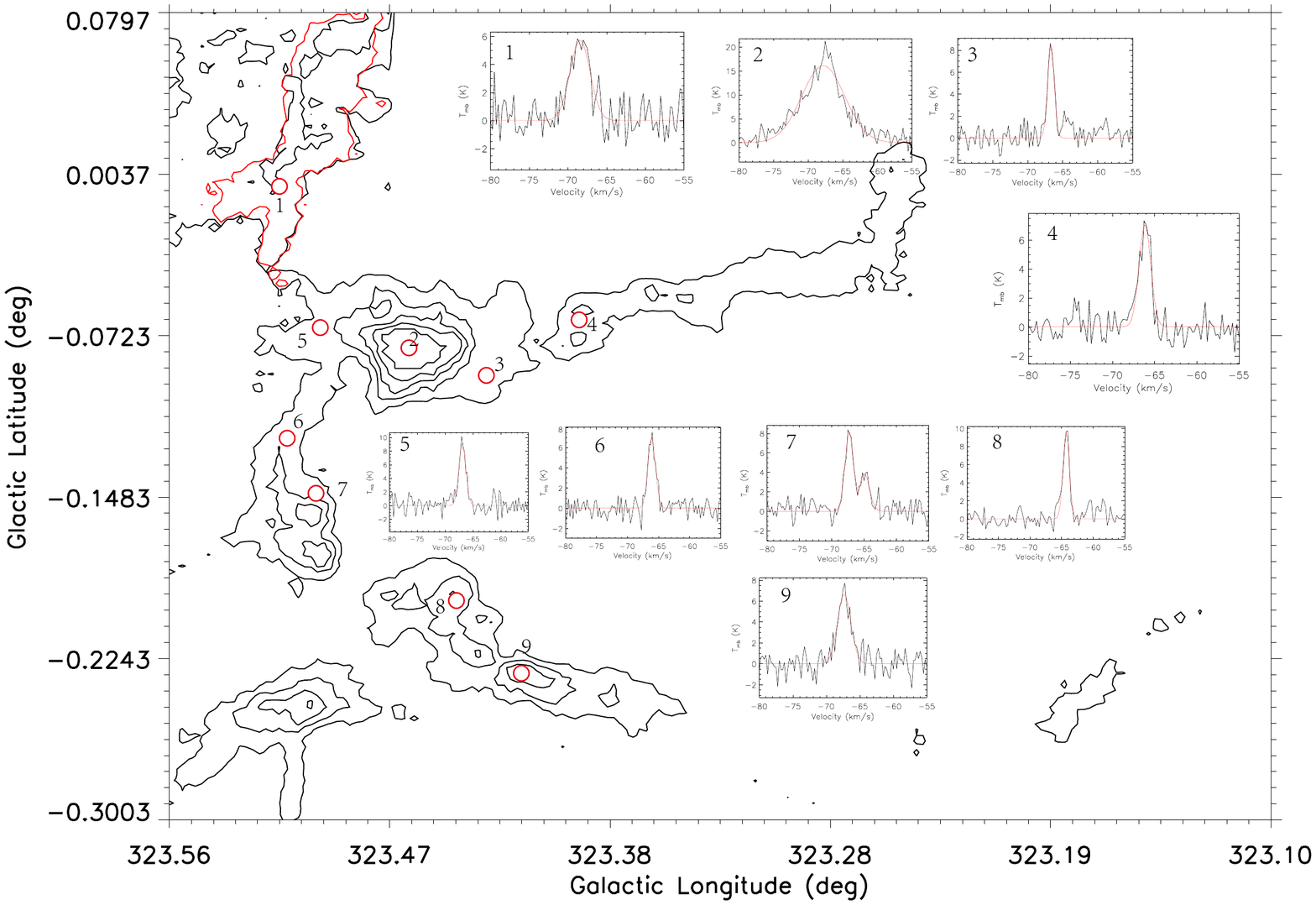}
  \caption{ Black contours are $^{13}$CO\,($J$\,=\,2--1) integrated intensity from velocity -69 to -63\,km\,s$^{-1}$ and the red contour is $^{13}$CO\,($J$\,=\,2--1) integrated intensity from velocity -69 to-68.75\,km\,s$^{-1}$. The spectral lines at the corresponding positions of the numbers are shown, the red lines are the Gaussian fitting results.}
         \label{fig_spectrums_appendix} 
  \end{figure*}
\end{appendix}
\end{document}